\documentstyle[aps,eqsecnum]{revtex}
\begin{document}

\newcommand{\be}{\begin{equation}}
\newcommand{\ee}{\end{equation}}
\newcommand{\bea}{\begin{eqnarray}}
\newcommand{\eea}{\end{eqnarray}}
\newcommand{\nn}{\nonumber}

\setcounter{page}{1}
\draft

\title{Unconstrained $SU(2)$ Yang-Mills Quantum Mechanics with Theta Angle}
\author{A M. Khvedelidze \ $^a$
\thanks{Permanent address: Tbilisi Mathematical Institute,
380093, Tbilisi, Georgia},
\,\, H.-P. Pavel\ $^b$ and \,\, G. R\"opke $^b$}
\address{$^a$ Bogoliubov Laboratory of Theoretical Physics , 
Joint Institute for Nuclear Research, Dubna, Russia}
\address{$^b$ Fachbereich Physik der Universit\"at Rostock,
              D-18051 Rostock, Germany}
\date{September 27, 1999}
\maketitle

\begin{abstract}
The unconstrained classical system equivalent to spatially homogeneous 
$SU(2)$ Yang-Mills theory with theta angle is obtained and canonically 
quantized. The Schr\"odinger eigenvalue problem is solved approximately 
for the low lying states using variational calculation. 
The properties of the groundstate are discussed, in particular its electric 
and magnetic properties, and the value of the "gluon condensate" is 
calculated. Furthermore it is shown that the energy spectrum of $SU(2)$ 
Yang-Mills quantum mechanics is independent of the theta angle.
Explicit evaluation of the Witten formula for the topological 
susceptibility gives strong support for the consistency of the 
variational results obtained.
\end{abstract}

\bigskip

\bigskip


\bigskip
\bigskip

\section{Introduction}

One of the central problems in the investigation  
of non-Abelian gauge theories is a gauge invariant description of
the vacuum and the low lying exited states.
In the standard approach to the quantization of gauge theories
the physical states have to satisfy not only the
Schr\"odinger equation but additionally be unnihilated by the Gauss
law operator to implement gauge invariance at the quantum level
\cite{Jackiw}. However it is well known that
there exist states which satisfy the Gauss law but are not
invariant under the so-called homotopically nontrivial gauge
transformations, leading to the appearence of the theta angle
\cite{JackiwRebbi},\cite{Callan}.
A well-elaborated semiclassical approach to the theta structure of the
groundstate has been given in the ``instanton picture", where
the theta angle is interpreted \cite{Jackiw} in analogy to the
Bloch momentum in Solid State Physics.
The instantons, which are selfdual solutions of the Euclidean classical
equations of motion with finite action,
correspond to semiclassical quantum mechanical tunneling paths in
Minkowski space between the infinite sequence of degenerate
zero-energy Yang-Mills vacua of different homotopy classes of the gauge
potential.
The semiclassical instanton picture of
the theta vacuum however is of course reliable only for weak coupling.
For a complete investigation of the theta stucture
of the vacuum of Yang-Mills quantum theory a rigorous treatment
at strong coupling is necessary.
The effect of the theta angle for arbitrary coupling constant
can be taken into account by adding the Pontryagin density to the
Yang-Mills Lagrangian \cite{Jackiw} .
Although the extra theta dependent CP-violating term is only a total 
divergence and therefore has no meaning classically, it can have a physical 
meaning at the quantum level as is still under lively discussion
\cite{Adam}-\cite{Gaba1}.

As a first step towards a full investigation of Yang-Mills theory in the
strong coupling limit the toy model of $SU(2)$ Yang-Mills mechanics of
spatially homogeneous fields has been considered
on the classical \cite{Matinyan} -\cite{GKMP} as well as on the 
quantum level \cite{Simon}-\cite{gymqm}.
In the present paper we will analyse the model of $SU(2)$ Yang-Mills 
mechanics of spatially homogeneous fields for arbitrary theta angle. 
In order to obtain the equivalent unconstrained classical system 
in terms of gauge invariant variables only \cite{GoldJack} -\cite{GKP}, 
we apply
the method of Hamiltonian reduction (\cite{GKP} and references therein)
in the framework of the  Dirac constraint formalism 
\cite{DiracL} -\cite{HenTeit}. As in our recent work \cite{GKMP}
the elimination of the pure gauge degrees of freedom is achieved
by using the polar representation for the gauge potential, 
which trivializes the Abelianization of the Gauss law constraints,
and finally projecting onto the constraint shell. The obtained
unconstrained system then describes the dynamics of a symmetrical second
rank tensor under spatial rotations. The main-axis-transformation of this
symmetric tensor allows us to separate the gauge invariant variables
into scalars under ordinary space rotations and into ``rotational'' degrees
of freedom. In this final form the physical Hamiltonian
and the topological operator can be quantized
without operator ordering ambiguities.
We study the residual symmetries of the resulting unconstrained
quantum theory with arbitrary theta angle and reduced the eigenvalue 
problem of the Hamiltonian to the 
corresponding problem with zero theta angle.
Using the variational approach we calculate 
the low energy spectrum with rather high accuracy.
In particular we find the energy eigenvalue 
and the magnetic and electric properties of the groundstate, 
as well the corresponding value of the "gluon condensate".
The groundstate energy is found to be independent of the
theta angle by construction of the explicit transformation
relating the Hamiltonians with different theta parameter.
This is confirmed by an explicit calculation of the Witten formula 
for the topological susceptibility using our variational results for the
groundstate and the low lying excitations give strong support
for the independence of the groundstate energy of theta thus
indicating the consistence of our results.

Our paper is organized as follows. In Section II the Hamiltonian reduction
of $SU(2)$ Yang-Mills mechanics for arbitrary theta angle
is carried out and the corresponding unconstrained system put into
a form where the rotational and the scalar degrees of freedom are maximally
separated. In Section III the obtained unconstrained classical
Hamiltonian is quantized, its residual symmetries, the necessary
boundary conditions for the wave functions
and the relevance of the theta angle on the quantum level discussed. 
In Section IV the eigenvalue problem of the unconstrained
Hamiltonian with vanishing theta angle is solved approximately
in the low energy region using the variational approach.
In Section V the Witten formula for the topological susceptibility
is evaluated using the obtained variational results.
Section VI finally gives our conclusions.
Appendices A to C state several results and additional discussions
relevant to the main text.

\section{Unconstrained classical $SU(2)$ Yang-Mills mechanics 
with theta angle}
\subsection{Hamiltonian formulation}

It is well known \cite{Jackiw} that the theta angle can be included
already at the level of the classical action
\be
\label{eq:act}
S [A] : = - \frac{1}{4}\ \int d^4x\ \left(F^a_{\mu\nu} F^{a\mu \nu}
- {\alpha_s\theta\over 2\pi} F^a_{\mu\nu} \tilde{F}^{a\mu \nu}\right)~,
\ee
with the $SU(2)$ Yang-Mills field strengths
$F^a_{\mu\nu} : = \partial_\mu A_\nu^a  -  \partial_\nu A_\mu^a
+ g \epsilon^{abc} A_\mu^b A_\nu^c~$, ($a=1,2,3$), the dual
$\tilde{F}_{a\mu \nu}:= {1/ 2} \epsilon_{\mu\nu\sigma\rho}
F^{a\sigma \rho}$ and $\alpha_s={g^2/  4\pi}$.
For the special case of spatially homogeneous fields the Lagrangian
in (\ref{eq:act}) reduces to
\footnote{ Everywhere in the paper we put the spatial volume $V= 1$.
 As result the coupling constant $g$ becomes dimensionful
 with $g^{2/3}$ having the dimension of energy. The volume dependence
 can be restored in the final results by replacing $g^2$ by $g^2/V$. }
\be \label{hl}
L={1\over 2}\left(\dot{A}_{ai}-g\epsilon_{abc}A_{b0} A_{ci}\right)^2
   -{1\over 2} B_{ai}^2  -{\alpha_s\theta\over 2\pi}
       \left(\dot{A}_{ai}-g\epsilon_{abc}A_{b0} A_{ci}\right)B_{ai}~,
\ee
with  the magnetic field
$B_{ai}= (1/2)g\epsilon_{abc}\epsilon_{ijk}A_{bj}A_{ck}$.
After the  supposition of spatial homogeneity of the fields
the $SU(2)$ gauge invariance of  the Yang-Mills  action 
Eq. (\ref{eq:act}) reduces to the symmetry under  the $SO(3)$ 
local transformations 
\bea
A_{a0}(t)& \longrightarrow & A^{\omega}_{a0}(t)=
O(\omega(t))_{ab}A_{b0}(t) -\frac{1}{2g} 
\epsilon_{abc}\left(O(\omega(t))\dot O(\omega(t)) \right)_{bc}\,,\nn\\
A_{ai}(t)& \longrightarrow & A^{\omega}_{ai}(t)=
O(\omega(t))_{ab}A_{bi}(t)\label{tr} 
\eea 
and as a result the  Lagrangian  (\ref{hl}) is degenerate.
From the calculation of the canonical momenta
\be
P_{a}  :=  \partial L/\partial (\partial_0{A}_{a0} ) = 0 \,,\,\,\,\,\,\,\,
\Pi_{ai} := \partial L/\partial (\partial_0 {A}_{ai})
          = \dot{A}_{ai}-g\epsilon_{abc}A_{b0} A_{ci}
            - {\alpha_s\theta\over 2\pi} B_{ai}~,
\ee
one finds that the phase space spanned by the variables
\( (A_{a0},  P_a) \) and  \( (A_{ai}, \Pi_{ai}) \)
is restricted by the  primary constraints
$P_a (x) = 0~$. The evolution of the system is governed
by the total Hamiltonian  \cite{DiracL} with three arbitrary functions
\(\lambda_a (x)\)
\be
\label{Htot}
H_T := \frac{1}{2}\Pi_{ai}^2
+ \frac{1}{2} \left(1+ \left({\alpha_s\theta\over 2\pi}\right)^2\right) 
B_{ai}^2 (A) +\theta Q(\Pi,A)
 - g A_{a0}  \epsilon_{abc}A_{bi}\Pi_{ci} +
\lambda_a (x) P_a (x)~,
\ee
where the topological charge has been introduced 
\be
Q := -{\alpha_s\over 2\pi}\Pi_{ai}B_{ai}~.
\ee
Apart  from  the primary constraints $P^a = 0~$
the phase space is restricted also by the 
non-Abelian Gauss law, the secondary constraints 
\be
\label{eq:secconstr}
\Phi_a : = g\epsilon_{abc} A_{ci} \Pi_{bi} = 0~,
\qquad
\{\Phi_i  , \Phi_j \} =  g\epsilon_{ijk}\Phi_k ~,
\ee
which follow from the maintenance  of the primary constraints in time.

To overcome the problems of the existence of these constraints and the 
nonunique character of the dynamics governed by the 
total Hamiltonian (\ref{Htot}) we will follow the method of 
Hamiltonian  reduction to constuct  the  unconstrained system 
with  uniquely predictable dynamics. 
As in the recent paper \cite{GKMP} we shall use a special set of 
of coordinates which is very suitable for the 
implementation of Gauss law constraints  and the 
derivation of the physically relevant theory 
equivalent to the initial degenerate theory.
This will be the subject of the following Subsection.

\subsection{Canonical transformation to adapted coordinates 
and projection to Gauss law  constraint}

The local symmetry transformation 
(\ref{tr}) of the gauge potentials 
\( A_{ai} \) promts us with the set of  coordinates in terms of which the 
separation of  the  gauge degrees of freedom occurs.
As in \cite{GKMP} we use the  polar decomposition  for arbitrary 
\(3\times 3\) quadratic matrices \cite{Marcus} 
\be
\label{eq:pcantr}
A_{ai} \left(\chi, S \right)
= O_{ak}\left( \chi \right) S_{ki}~,
\ee
with the orthogonal matrix \( O (\chi)  \),
parametrized by the three angles \(\chi_i\)  and the positive definite 
\(3\times 3\) symmetric matrix \( S \).
The representation (\ref{eq:pcantr}) can be regarded as transformation  
from the gauge potentials \( A_{ai}\) to 
a the set of coordinates \(\chi_i\)  and \( S_{ik} \).
The corresponding canonical conjugate momenta \( (p_{\chi_i}, P_{ik} )\)
can be obtained using the generating function
\be
{F} \left( \Pi; \chi, S \right)=
\sum_{a,i}^3 \Pi_{ai} A_{ai} \left(\chi, S\right) =
\mbox{tr}
\left( \Pi^T O (\chi) S  \right)
\ee
as
\bea
p_{\chi_j} & = & \frac{\partial F}{\partial \chi_j} =
\sum_{a,s,i}^3 \Pi_{ai} \ \frac{\partial O_{as}}{\partial \chi_j } \
S_{ si}  =
\mbox{tr}
\left[ \Pi^T \frac{\partial O}{\partial \chi_j} \ S
\right],\\
 P_{ik}  & = & \frac{\partial F}{\partial S_{ik}} =
\frac{1}{2} \left( O \Pi^T  + \Pi O^T  \right)_{ik}~.
\eea
A straightforward calculation \cite{GKMP} yields the following expressions
for the field strengths \( \Pi_{ai} \)
in terms of the new canonical variables
\be  \label{eq:potn}
\Pi_{ai} = O_{ak}(\chi) \biggl[{\,  P_{ki} +
\epsilon _{kli}
 ( s^{-1})_{lj} \left[
\left(\Omega^{-1}(\chi) p_{\chi} \right)_{j}  - \epsilon_{mjn} \left(
PS\right)_{mn}\,\right]\,
}\biggr]~,
\ee
with
\be
\Omega_{ij}(\chi) \, : = \,\frac{1}{2} \, \epsilon_{min}
\left[ \frac{\partial O^T \left(\chi\right)}{\partial \chi_j}
\, O (\chi) \right]_{mn}~,
\ee
and
\be
s_{ik} : = S_{ik} -  \delta_{ik}
\mbox{tr} S~.
\ee
Using the representations (\ref{eq:pcantr}) and (\ref{eq:potn})
one can easily convince oneself that the
variables \( S \) and \(P \) make no contribution to the
Gauss law constraints (\ref{eq:secconstr})
\be
\Phi_a : = O_{as}(\chi) \Omega^{-1}_{\ sj}(\chi) p_{\chi_j} = 0~.
\label{eq:4.54}
\ee
Hence, assuming the invertibility of  the matrix 
$ \Omega$, the non-Abelian Gauss law  constraints  are  
equivalent to  the set of Abelian  constraints 
\be
p_{\chi_a}   = 0~.
\ee

After having rewritten the model in terms of 
adapted canonical pairs and after Abelianization of the Gauss 
law constraints (\ref{eq:secconstr}) the construction of 
the unconstrained Hamiltonian system
can be obtained as follows. 
The  physical unconstrained Hamiltonian, defined  as 
\[
H_{\theta}(S,P):= H_T( S, P) \,
 \Bigl\vert_{ p_{\chi_a}= 0}~,
\]
takes the form 
\be \label{eq:uncYME}
H_{\theta} \, = \,
\frac{1}{2}  \mbox{tr}({\cal{E}}^2)
+ \frac{g^2}{4} (1+{\alpha_s^2\over 4\pi^2}\theta^2) \left[
\mbox{tr}{}^2 ( S )^2  -
 \mbox{tr} (S)^4
\right]
+ \theta Q(S,P) \, ,
\ee
where the  ``physical'' electric field strengths
\( {\cal{E}}_{ai} \) are 
\be
\Pi_{ai}\ \biggl\vert_{\pi_a = 0} =: \ O_{ak}(q) \
{\cal{E}}_{ki}
(S, P)~,
\ee
and the topological charge
\be
Q(S,P) = -{\alpha_s\over 2\pi}\mbox{tr}\left(PS\right)~.
\ee
Using  the representation (\ref{eq:potn})
for the electric field  one can  express  the \( {\cal{E}}_{ai} \)
in terms of  the physical variables \( P \) and \( S \)
\be  \label{eq:els}
{\cal{E}}_{ki}(S , P) =
P_{ik} +  \frac{1}{\det s}
 \left( s {\cal M} s \right)_{ik}\,.
\ee
where  \({\cal M}\) denotes the spin part of the angular angular 
momentum tensor of the initial gauge field 
\be
{\cal M}_{mn}:= \left( S P - PS \right)_{mn}~.
\ee
Using (\ref{eq:els}) the  unconstrained Yang-Mills Hamiltonian reads
\be \label{eq:uncYMP}
H_{\theta} \left(S,P\right) =
\frac{1}{2}  \mbox{tr}(P)^2  +
\frac{1}{2 \det^2 s }
\mbox{tr}\, \left(s {\cal M} s \right)^2
+ \frac{g^2}{4}\left(1+{\alpha_s^2\over 4\pi^2}\theta^2\right) \left[
\mbox{tr}{}^2 (S)^2  -   \mbox{tr} (S)^4
\right] + \theta Q(S,P) \, .
\ee

\subsection{Unconstrained Hamiltonian in terms of 
rotational and scalar degrees of freedom}

In order to achieve a more transparent form for the reduced Yang-Mills
system (\ref{eq:uncYMP}) it is convenient  
to decompose the positive definite symmetric matrix \( S\) as
\be
S  =  R^{T}(\alpha,\beta,\gamma)\  D (x_1,x_2,x_3) \ 
R(\alpha,\beta,\gamma)~,
\ee
with the \( SO(3)\) matrix  \({R}\)
parametrized by the three Euler angles \((\alpha,\beta,\gamma )\),
and the diagonal matrix
\be
 D : = \mbox{diag}\ ( x_1 , x_2 , x_3 )~.
\ee
Using the $x_i$ and the Euler angles
\((\alpha,\beta,\gamma )\) and the corresponding canonical momenta $p_i$
and $p_\alpha,p_\beta, p_\gamma $ 
as the new set of canonical variables on the unconstrained phase space
we get the following physical Hamiltonian 
\be
\label{eq:PYM}
H_{\theta}\left(x_i,p_i;\xi_i\right) = \frac{1}{2} \sum_{cyclic}^3  
\left[
p^2_i + \xi^2_i \frac{x_j^2 +  x_k^2}{\left(x_j^2 -
x_k^2\right)^2} +
     g^2 (1+{\alpha_s^2\over 4\pi^2}\theta^2) \
x_j^2 x_k^2 \right] + \theta Q(p,x)~.
\ee
In (\ref{eq:PYM}) all  rotational variables are   
are combined into the quantities   $\xi_i$  
\bea
&& \xi_1  : =
\frac{\sin\gamma}{\sin\beta}\ p_\alpha +
\cos\gamma \  p_\beta - \sin\gamma \cot\beta \ p_\gamma~,\\
&& \xi_2 : =
-\frac{\cos\gamma}{\sin\beta}\ p_\alpha +
\sin\gamma \  p_\beta + \cos\gamma \cot\beta \ p_\gamma~,   \\
&& \xi_3  : = p_\gamma~.
\eea
representing  the $SO(3)$ invariant  Killing vectors with the Poisson 
brackets algebra  
\be
\{\xi_i,\xi_j\}=-\epsilon_{ijk}\xi_k~.
\ee
The topological charge $Q$ is independent of the rotational
degrees of freedom and depends on the diagonal 
canonical pairs in the particularly simple cyclic form
\be
Q 
=-g{\alpha_s\over 2\pi}\left(x_1 x_2 p_3 + x_2 x_3 p_1 + x_3 x_1 p_2 
\right)~.
\ee

This completes our reduction of the spatially homogeneous constrained
Yang-Mills system with theta angle to the equivalent 
unconstrained  system
describing the dynamics of the physical degrees of freedom. 

If we would restrict our consideration  only to the classical level,
the above generalization to arbitrary theta angle would be unnecessary,  
because the  theta dependence 
enters the initial Lagrangian  in the form of a total time derivative 
and thus the  value  of the theta angle has no influence  
on the classical equations of motion.
In the Hamiltonian formulation one can easily verify that the theta 
dependence can be removed from the Hamiltonian $H_{\rm \theta}$ by 
the  canonical  transformation to the new variables  
\bea
\tilde{p}_i &:=& p_i - g{\alpha_s\theta\over 2\pi} x_j x_k\,, 
\qquad  i,j,k\ {\rm cyclic}~,\nonumber\\
\tilde{x}_i &:=& x_i~.
\label{clctr}
\eea
However, the transition to the quantum level requires a more careful 
treatment of the problem. It is necessary to clarify 
whether the operator corresponding to (\ref{clctr})
acting on the  quantum states is unitary. 

In subsequent Sections we shall consider the quantum treatment  of 
the obtained classical system and shall 
discuss the theta dependence of the vacuum in this model. 
    
\section{Quantization, symmetries and boundary conditions}

The Hamilton  operator  corresponding  to (\ref{eq:PYM}) is obtained
in the Schr\"odinger  configuration representation  by the
conventional  representation  for the canonical 
momenta \(p_k= -i \partial/\partial x_k\)
\be
\label{HHHq}
H_\theta:={1\over 2}\sum_{\rm cyclic}^3
\left[-{\partial^2\over\partial x_i^2} + \xi^2_i
{x_j^2+x_k^2 \over (x_j^2-x_k^2)^2}
+
g^2\left(1+{\alpha_s^2\over 4\pi^2}\theta^2
\right)x_j^2 x_k^2~\right] + \theta Q~,
\ee
with the topological charge operator 
\be
Q=  ig\frac{\alpha_s}{2\pi} 
\sum_{\rm cyclic}^3{x_ix_j}{\partial\over\partial x_k}~,
\ee
and the intrinsic angular momenta  $\xi$
obeying the commutation relations
\be
[\xi_i,\xi_j]=-i\epsilon_{ijk}\xi_k~.
\ee
The transition  to the quantum system 
in this adapted basis is free from operator ordering ambiguities.

As already mentioned in the last section  the parameter 
theta is unphysical on the classical level, since it can be
removed from Hamiltonian $H_{\rm \theta}$ by the  canonical 
transformation (\ref{clctr}).  
One can easily convince oneself that the quantum Hamiltonians 
$H_{\rm \theta}$ and  $H_{\rm \theta=0}$ can be related to each other
via the transformation
\be
\label{UHU1}
H_{\rm \theta} = U(\theta) H_{\rm \theta=0} U^{-1}(\theta)~,
\ee
with
\be
\label{UHU2}
U(\theta)= \exp[ig{\alpha_s\over 2\pi}\theta x_1 x_2 x_3]~.
\ee
The question is whether this operator is unitary in the 
domain of definition of the Hamiltonians $H_{\rm \theta}$ 
and  $H_{\rm \theta=0}$, 
which is determined by their respective symmetries and the boundary 
conditions to be imposed on the corresponding wave functions.

\subsection{ Boundary conditions}

Due to the positivity of the coordinates $x_i$ in the polar decomposition 
(\ref{eq:pcantr}) the configuration space is $R^{+}_3$
after the elimination of the pure gauge degrees of freedom.   
Thus the implementation of the canonical rules of quantization to the 
unconstrained classical system requires the specification of the 
boundary conditions both at positive infinity and on the three 
boundary planes $x_i=0~,\ \ i=1,2,3$.
The requirement of Hermiticity of the Hamiltonian $H_{\rm \theta}$ 
(\ref{HHHq}) leads to the condition
\be
\label{bctheta}
\left(\Psi^*_\theta\partial_k\Phi_\theta-\partial_k\Psi^*_\theta\Phi_\theta
+2ig{\alpha_s\over 2\pi}\theta x_i x_j\Psi^*_\theta\Phi_\theta
\right)\Big|_{x_k=0}=0~,\ \ \ \ i,j,k\ {\rm cyclic}~.
\ee
Using the relation $\Psi_\theta=U(\theta)\Psi_{\theta=0}$ with $U(\theta)$
given in (\ref{UHU2}), this reduces to the corresponding requirement
for the Hermiticity of $H_{\rm \theta=0}$
\be
\label{bctheta0}
\left(\Psi^*_{\theta=0}\partial_k\Phi_{\theta=0}
-\partial_k\Psi^*_{\theta=0}\Phi_{\theta=0}
\right)\Big|_{x_k=0}=0~,\ \ \ \ i,j,k\ {\rm cyclic}.
\ee
It is satisfied for ($\kappa$ arbitrary c-number)
\be
\label{bctheta0e}
\left(\partial_k\Psi_{\theta=0}+\kappa\Psi_{\theta=0}\right)
\Big|_{x_k=0}=0~,\ \ \ \ k=1,2,3~,
\ee
which includes the two limiting cases of vanishing wave function 
($\kappa\to\infty$) or vanishing derivative of the wave function 
($\kappa=0$) at the boundary.
The requirement of the Hermiticity of the momentum operators
in  the  Schr\"odinger configuration representation  
$ p_i := -i\partial/ \partial x_i $ on $R^{+}_3$
requires the wave function to obey the boundary conditions 
\bea
\label{bc1}
&&\Psi_{\theta=0}\Big|_{x_i=0}= 0~, \ \ \ \ \ \ \ i=1,2,3~,\\
\label{bc2}
&&\Psi_{\theta=0}\Big|_{x_i \to \infty} =0~,\ \ \ \ \ \ \ i=1,2,3~.
\eea
In particular, they also imply the Hermiticity and the existence of a real
eigenspectrum of the topological charge operator $Q$.
Its eigenstates, however, given explicitly in Appendix A, do not satisfy
the boundary conditions (\ref{bc1}) and (\ref{bc2}), similar to the
eigenstates of the momentum operator $-i\partial/ \partial x_i $.
Furthermore, it is interesting to note  that the characteristics 
of the $Q$ operator coincide with the Euclidean self(anti-)dual zero-energy
solutions of the classical equations of motion. They are the
analogs of the instanton solutions, but do not correspond to
quantum tunnelling between different vacua (see Appendix A).

\subsection{Symmetries of the Hamiltonians $H_{\rm \theta}$ 
and  $H_{\rm \theta=0}$}

As a relict of the rotational invariance of the initial gauge field theory
the Hamiltonian (\ref{HHHq}) possesses the symmetry
\be
\label{cHI}
[H,J_k]=0~,
\ee
where $J_i= R_{ij}\xi_j$ are the spin part of the generators of the angular 
momentum of Yang-Mills fields satisfying the $so(3)$ algebra
\be
[J_i, J_j] = i \epsilon_{ijk} J_k~, 
\ee
and commuting with the intrinsic angular momenta, $[J_i, \xi_j] = 0~$.
Hence the eigenstates can be classified 
according to  the quantum numbers $J$ and $M$ as the eigenvalues 
of the spin  $\vec{J}^2= J_1^2 + J_2^2+ J_3^2 $ and $J_3$.
The Hilbert spaces of states with different spin $J$
are each invariant subspaces under the action of 
all generators \(J_i\) and can therefore be considered
as separate eigenvalue problems.

Apart from this continous rotational symmetry the Hamiltonians
$H_{\rm \theta}$ and  $H_{\rm \theta=0}$ possess the following
discrete symmetries.
Both $H_{\rm \theta=0}$ and $Q$
are invariant under arbitrary permutations of any two of the 
variables $\sigma_{ij}x_i = x_j\sigma_{ij}, \, 
\sigma_{ij}p_i =p_j\sigma_{ij}$ 
\bea
[H_{\theta=0},\sigma_{ij}]=0~,\ \ \ \ \ \ \ [Q,\sigma_{ij}]=0~.
\eea
However, under time reflections 
$Tx_i = x_iT,\,\,Tp_i = - p_iT$, as well as 
under parity reflections ${\cal P} x_i= -x_i {\cal P},\,
{\cal P} p_i= -p_i {\cal P}$,
$H_{\rm \theta=0}$ commutes with $T$ and ${\cal P}$,
\be
[H_{\rm \theta=0},T]=0~,\ \ \ \ \ \ \ [H_{\theta=0},{\cal P}]=0~,
\ee
but $Q$ anticommutes with $T$ and ${\cal P}$,
\be
QT =-TQ~,\ \ \ \ \ \ \ 
Q{\cal P} =-{\cal P}Q~.
\ee
Hence for the $H_{\rm \theta=0}$ Schr\"odinger eigenvalue problem we can
restrict to the Hilbert space of real and parity odd wave functions
which automatically satisfy the boundary conditions (\ref{bc1}).
Observe that the transformation (\ref{UHU2}) leads out of the
corresponding Hilbert space and is therefore not unitary. 

\subsection{Independence of the energy spectrum of the theta angle}

Due to the relation (\ref{UHU1}) between the Hamiltonians $H_{\rm \theta}$,
and  $H_{\rm \theta=0}$ and the corresponding compatibility of the boundary
conditions discussed above the energy spectrum
should be independent of the theta angle.   
In particular the topological susceptibility of the vacuum should vanish.
Using the Witten formula \cite{Witten},\cite{Diakonov}, the 
topological susceptibility can be represented as the sum of a propagator
term involving the transition matrix elements of the topological
operator $Q$ and a contact term prportional to the vacuum expectation
value of the square of the magnetic field.
Independence of the groundstate energy of the theta angle
and hence vanishing topological susceptibility should therefore imply
\be
\label{WF0}
{d^2 E_0(\theta)\over d\theta^2}\Big|_{\theta = 0}=
-2\sum_n{|\langle 0|Q|n\rangle |^2\over E_n-E_0}
+ \langle 0|\left({\alpha_s\over 2\pi}\right)^2 B^2|0 \rangle = 0~,
\ee
where $|n>$ are eigenstates of the Hamiltonian $H_{\rm \theta=0}$ 
with energy eigenvalues $E_n$.
As we shall see below that our calculation of the
low energy part of the spectrum of $H_{\rm \theta=0}$ using the 
variational technique is in full accordance with (\ref{WF0}).

\section{Schr\"odinger eigenvalue problem for vanishing theta} 

\subsection{Low energy spin-0 spectrum from variational calculation}

The Hilbert space of states with zero spin $\vec{J}^2=0$
is an invariant subspace under the action of 
all generators \(J_i\) and one can consider the eigenvalue 
problem separately from states characterized by higher 
spin value.
Thus in the sector of zero spin $\vec{J}^2=\vec{\xi}^2=0$ the 
Schr\"odinger eigenvalue problem (\ref{HHHq}) reduces to  
\be
\label{H--0}
 H_0 \Psi_E \equiv 
{1\over 2}\sum_{\rm cyclic}^3\left[-{\partial^2\over\partial x_i^2}
+ g^2 x_j^2 x_k^2\right]\Psi_E = E \Psi_E~.
\ee
We shall use the boundary conditions (\ref{bc1}) and (\ref{bc2}).
Already a long time ago it has been proven by F. Rellich \cite{Rellich} 
that Hamiltonians of the type (\ref{H--0}) have a 
discrete spectrum due to quantum fluctuations, although the classical problem 
allows for scattering trajectories (see discussion in \cite{Simon}).
Related and simplified versions of the  eigenvalue problem (\ref{H--0}) 
have been studied extensively by many authors 
using different methods \cite{Simon}-\cite{Gaba2}.
In particular, in \cite{Medvedev}-\cite{BartBrunRaabe} the eigenstates and 
eigenvalues have been found in the semiclassical approximation for the 
special two dimensional case $x_3=0$.
\footnote{ It is interesting that for the  three dimensional case
one can write the potential term  in 
(\ref{H--0}) in the form
$V=\sum_{i=1}^3 (\partial_i W)^2$
with the "superpotential" \(~W(x_1, x_2, x_3)=x_1x_2x_3 \).
Note that in the simplified two dimensional case
there is no such superpotential. The two-dimensional superpotential 
$W^{(2)}=xy$ corresponds to the two-dimensional harmonic oscillator 
$V^{(2)}=x^2+y^2$.
From the form of the superpotential it follows that the wave function 
$\Psi_0 =\exp[-gW]$
solves the Schr\"odinger eigenvalue problem with energy eigenvalue $E=0$.
It is the unconstrained, strong coupling form of the well-known exact but
nonnormalizable zero-energy solution \cite{Loos} of the Schr\"odinger
equation of Yang-Mills field theory .
Obviously it is also not satisfying the boundary conditions (\ref{bc1}), 
(\ref{bc2}) and has to be disregarded as a false groundstate.}

To obtain the approximate low energy spectrum of the Hamiltonian
in the spin-0 sector 
we will use the Rayleigh-Ritz variational method \cite{ReedSimon}
based on the minimization of the energy functional 
\be
\label{energyf}
{\cal E}[\Psi] := \frac{\big<\Psi|H_0|\Psi\big>}
{\big<\Psi|\Psi\big>}~.
\ee 
The key moment in all variational calculations 
is the choice of the trial functions. 
Guided by the harmonic oscillator form of the valleys of the potential
in (\ref{H--0}) close to the bottom a simple first choice for a trial 
function compatible  with the boundary conditions (\ref{bc1}) and 
(\ref{bc2}) is to use \cite{gymqm} the lowest state of three 
harmonic quantum oscillators on the positive half line      
\be
\label{Psi000}
\Psi_{000}=
8\prod_{i=1}^3 \left({\omega_i\over \pi}\right)^{1/4}\sqrt{\omega_i}x_i
e^{-\omega_ix_i^2/2}~.
\ee
The stationarity conditions for the energy functional 
of this state,   
\bea
{\cal E}[\Psi_{000}]=
\sum_{cyclic}^3 \left({3\over 4}\omega_i+
       {9\over 8}g^2{1\over \omega_j\omega_k}\right)~,\nn
\eea
lead to the isotropic optimal choice  
\be \label{fr}
\omega :=\omega_1=\omega_2=\omega_3=3^{1/3}g^{2/3}~.
\ee
As a first upper bound for the groundstate energy of the Hamiltonian we 
therefore find
\bea
\label{1stest}
E_0 \le {\cal E}[\Psi_{000}]=
{27\over 8}3^{1/3}g^{2/3} = 4.8676~ g^{2/3}.
\eea
The upper bound (\ref{1stest}) is in agreement with the lower bound of 
the energy functional for separable functions   
\be 
{\cal E}[\Psi_{\rm sep}] \ge  4.5962~ g^{2/3}~,
\ee
derived in Appendix B.

In order to improve the upper bound for the groundstate energy
of the Hamiltonian $H_0$ we extend the space of trial functions 
(\ref{Psi000}) 
and consider the Fock space of the orthonormal set of products 
\be 
\label{bel}
\Psi_{n_1 n_2 n_3}:=
\prod_{i=1}^3 \Psi_{n_i}(\omega, x_i)~,
\ee
of the odd eigenfunctions of the harmonic oscillator
\bea
\Psi_{n}(\omega,x):=
            {(\omega/\pi)^{1/4}\over \sqrt{2^{2n}(2n+1)!}}
e^{-\omega x^2/2}
            H_{2n+1}(\sqrt{\omega}x)~,\nn
\eea
with the frequency fixed by (\ref{fr}).

Furthermore the variational procedure becomes much more effective,
if the space of trial functions is decomposed into the irreducible
representations of the residual discrete symmetries of the Hamiltonian 
(\ref{H--0}). As has been discussed in Section III B, it is invariant under 
arbitrary permutations of any two of the 
variables $\sigma_{ij}x_i = x_j\sigma_{ij}, \, 
\sigma_{ij}p_i =p_j\sigma_{ij}$ and 
under time reflections $Tx_i = x_iT,\,\,Tp_i = - p_iT$, 
\bea
[H_0,\sigma_{ij}]=0,\,\,\,\,\qquad [H_0,T]=0\,.\nn
\eea
We shall represent these by the permutation operator $\sigma_{12}$, 
the cyclic permutation operator $\sigma_{123}$ and the time reflection
operator $T$, whose action on the states is 
\bea
\sigma_{123}\Psi(x_1,x_2,x_3)&=&\Psi(x_2,x_3,x_1)~,\nn\\ 
\sigma_{12}\Psi(x_1,x_2,x_3)&=&\Psi(x_2,x_1,x_3)~,\nn\\
T\Psi(x_1,x_2,x_3)&=&\Psi^*(x_1,x_2,x_3)~,\nn
\eea
and decompose the Fock space spanned by the functions (\ref{bel})
into the irreducible representations of the 
permutation group and time reflection $T$.
For given $(n_1,n_2,n_3)$ we define
\bea
\label{typeI}
\Psi_{nnn}^{(0)+}:=\Psi_{n n n}~,
\eea
if all three indices are equal (type I), the three states $(m=-1,0,1)$
\bea
\label{typeII}
\Psi^{(m)+}_{nns}:=
{1\over\sqrt{3}}\sum_{k=0}^2 e^{-2km\pi i/3}
\left(\sigma_{123}\right)^k\Psi_{n n s}~,
\eea
when two indices are equal (type II), and the two sets of three states
$(m=-1,0,1)$
\be
\label{typeIII}
\Psi^{(m)\pm}_{n_1n_2n_3}:=
 {1\over\sqrt{6}}\sum_{k=0}^2 e^{-2km\pi i/3}\left(\sigma_{123}\right)^k
 \left(1\pm \sigma_{12}\right)\Psi_{n_1 n_2 n_3}~,
\ee
if all $(n_1,n_2,n_3)$ are different (type III).
In this new orthonormal set of irreducible basis states 
$\Psi^{(m)\alpha}_{\bf N}$, 
the Fock representation of the Hamiltonian $H_0$ reads
\bea
\label{H_0irr}
H_0=\sum |\Psi^{(m)\alpha}_{\bf M}\rangle
\langle\Psi^{(m)\alpha}_{{\bf M}}|H_0|\Psi^{(m)\alpha}_{\bf N}\rangle
\langle\Psi^{(m)\alpha}_{\bf N}|~.\nn
\eea
The basis states $\Psi^{(m)\alpha}_{\bf N}$ are eigenfunctions of
$\sigma_{123}$ and $\sigma_{12}T$ 
\bea
\sigma_{123}\Psi^{(m)\pm}_{\bf N}
&=&e^{2m\pi i/3}\Psi^{(m)\pm}_{\bf N}~,\nn\\ 
\sigma_{12}T\Psi^{(m)\pm}_{\bf N}
&=&\pm \Psi^{(m)\pm}_{\bf N}~.
\eea
Under $\sigma_{12}$ and $T$ separately, however, 
they transform into each other
\bea
\sigma_{12}\Psi^{(m)\pm}_{\bf N}&=&\pm \Psi^{(-m)\pm}_{\bf N}~,\nn\\
T\Psi^{(m)\pm}_{\bf N}&=& \Psi^{(-m)\pm}_{\bf N}~.\nn
\eea
We therefore have the following irreducible representations. 
The singlet states $\Psi^{(0)+}$, the ``axial'' singlet states 
$\Psi^{(0)-}$, the doublets $(\Psi^{(+1)+};\Psi^{(-1)+})$ and 
the ``axial'' doublets $(\Psi^{(+1)-};\Psi^{(-1)-})$. 
Since the partner states of the doublets transform into each 
other under the symmetry operations $\sigma_{12}$ or $T$, the corresponding
values of the energy functional are equal.

The energy matrix elements of the irreducible states can then be expressed
in terms of the basic matrix elements as given in Appendix C.
Due to this decomposition of the Fock space into the irreducible
sectors, the variational approach allows us to give
upper bounds for states in each sector.  
The values of the energy functional for the states in each irreducible
sector with the smallest number of knots  
${\cal E}[\Psi_{000}^{(0)+}]= 4.8676~ g^{2/3}$,
${\cal E}[\Psi_{100}^{(\pm 1)+}]= 7.1915~  g^{2/3}$,
${\cal E}[\Psi_{012}^{(0)-}]= 13.8817~ g^{2/3}$, and
${\cal E}[\Psi_{012}^{(\pm 1)-}]= 15.6845~ g^{2/3}$
give first upper bounds for the lowest energy eigenvalues 
of the singlet, the doublet, the axial singlet, and the 
axial doublet states.

In order to improve the upper bounds for each irreducible sector, 
we truncate the Fock space at a certain number of knots 
of the wave functions and search for the corresponding states
in the truncated space with the lowest value of the energy functional.
We achieve this by diagonalizing 
the corresponding truncated Hamiltonian $H_{\rm trunk}$ to
find its eigenvalues and eigenstates. Due to the orthogonality of the
truncated space to the remaining part of Fock space the value of the energy
functional (\ref{energyf}) for the  eigenvectors of $H_{\rm trunk}$
coincides with the $H_{\rm trunk}$ eigenvalues.  

Including all states in the singlet sector with up to $5$ knots 
we find rapid convergence to the following energy expectation values for 
the three lowest states 
$S_1,S_2,S_3$ 
\bea
{\cal E}[S_1] &=& 4.8067~ g^{2/3}\ (4.8070~ g^{2/3}) ,\nonumber\\
{\cal E}[S_2] &=& 8.2515~ g^{2/3}\ (8.2639~ g^{2/3}) ,\nonumber\\
{\cal E}[S_3] &=& 9.5735~ g^{2/3}\ (9.6298~ g^{2/3}) ,
\eea
where the numbers in brackets show the corresponding result when
including only states up to $4$ knots into the variational calculation.
The lowest state $S_1$, given explicitly as 
\bea
\label{groundstate}
S_1 &=& 0.9946~ \Psi_{000}^{(0)+} + 0.0253~ \Psi_{001}^{(0)+}
    - 0.0217~ \Psi_{002}^{(0)+} - 0.0970~ \Psi_{110}^{(0)+}\nonumber\\
&& - 0.0005~ \Psi_{003}^{(0)+} - 0.0033~ \Psi_{012}^{(0)+} 
- 0.0146~ \Psi_{111}^{(0)+} - 0.0005~ \Psi_{004}^{(0)+}\nonumber\\
&& + 0.0040~ \Psi_{013}^{(0)+} - 0.0080 ~\Psi_{220}^{(0)+}
- 0.0038~  \Psi_{112}^{(0)+} + 0.0001~ \Psi_{005}^{(0)+}\nonumber\\
&& - 0.0004~ \Psi_{014}^{(0)+} + 0.0011~ \Psi_{023}^{(0)+}
   -0.0004~  \Psi_{113}^{(0)+} + 0.0031~ \Psi_{221}^{(0)+}~, 
\eea
nearly coincides with the state $\Psi_{000}^{(0)+}$, the contributions
of the other states are quite small.
Similarly including all states in the doublet sector with up to $6(5)$ 
knots the
following energy expectation values for the three lowest states 
$D_1^{(\pm 1)},D_2^{(\pm 1)},D_3^{(\pm 1)}$
\bea
{\cal E}[D_1^{(\pm 1)}] &=& 7.1682~ g^{2/3}\ (7.1689~ g^{2/3}) ,\nonumber\\
{\cal E}[D_2^{(\pm 1)}] &=& 9.6171~ g^{2/3}\ (9.6394~ g^{2/3}) ,\nonumber\\
{\cal E}[D_3^{(\pm 1)}] &=& 10.9903~ g^{2/3}\ (10.9951~ g^{2/3}).
\eea
have been obtained.
Including all states in the axial singlet sector with up to $8(7)$ knots 
we find the following energy expectation values for the three lowest 
states $A_1,A_2,A_3$
\bea
{\cal E}[A_1] &=& 13.2235~ g^{2/3}\ (13.2275~ g^{2/3}),\nonumber\\
{\cal E}[A_2] &=& 16.6652~ g^{2/3}\ (16.7333~ g^{2/3}),\nonumber\\
{\cal E}[A_3] &=& 19.1470~ g^{2/3}\ (19.3028~ g^{2/3}).
\eea
Finally taking into account all states in the axial doublet sector with up 
to $8(7)$ knots 
we find the following energy expectation values for the three lowest states 
$C_1^{(\pm 1)},C_2^{(\pm 1)},C_3^{(\pm 1)}$
\bea
{\cal E}[C_1^{(\pm 1)}] &=& 14.8768~ g^{2/3}\ (14.8796~ g^{2/3}),\nonumber\\
{\cal E}[C_2^{(\pm 1)}] &=& 17.6648~ g^{2/3}\ (17.6839~ g^{2/3}),\nonumber\\
{\cal E}[C_3^{(\pm 1)}] &=& 19.9019~ g^{2/3}\ (19.9914~ g^{2/3}),
\eea
We therefore obtain rather good estimates for the energies of the lowest
states in the spin-0 sector. Extending to higher and higher numbers of
knots in each sector we should be able to obtain the low energy spectrum
in the spin-zero sector to arbitrarily high numerical accuracy.

In summary comparing our results for the first few states in all sectors,
we find that the lowest state appears in the singlet sector with energy
\be 
\label{groundsten}
E_{0} = 4.8067~ g^{2/3}~,
\ee
with expected accuracy up to three digits after the dot.
Its explicit form is given in (\ref{groundstate}) to the accuracy 
considered.
For comparison with other work we remark that due to our boundary 
condition (\ref{bc1})
all our spin-0 states correspond to the $0^-$ sector in the work of
\cite{Martin} where a different gauge invariant representation of 
Yang-Mills mechanics has been used. 
Their state of lowest energy in this sector is $9.52~ g^{2/3}$. 
Furthermore in \cite{Gaba2}, using an analogy of $SU(N)$
Yang-Mills quantum mechanics in the large $N$ limit to membrane theory,
obtain the energy values $6.4690~ g^{2/3}$ and $19.8253~ g^{2/3}$ for
the groundstate and the first excited state.

The expectation values for the squares of the electric and the magnetic
fields for the groundstate (\ref{groundstate}) are found 
to be
\be
\langle 0|E^2|0\rangle = 6.4234~g^{2/3},\ \ \ \ \ \ \ \ 
\langle 0|B^2|0\rangle =3.1900~g^{2/3} ~,
\ee 
and the value for the "gluon condensate" is therefore
\be
\langle 0|G^2|0\rangle:= 2\left(\langle 0|B^2|0\rangle 
-\langle 0|E^2|0\rangle \right) = - 6.4669 ~g^{2/3}~.
\ee 
These results are expected to be accurate up to three digits after the dot.
Hence the variational calculation shows
that the vacuum is not self(anti-)dual and that a nonperturbative
"gluon condensate" appears.
\bigskip

\subsection{Higher spin states}

For the discussion of the eigenstates of the Hamiltonian $H_{\theta =0}$
with arbitrary spin we write
\be
\label{HHHs}
H_{\theta =0} = H_0 + H_{\rm spin}
\ee
with the spin-0 Hamiltonian (\ref{H--0}) discussed in the last 
subsection and the spin dependent part
\be
H_{\rm spin} = {1\over 2}\sum_{i=1}^3\xi_i^2 V_i~,\ \ \ \ 
V_i := {x_j^2+x_k^2\over (x_j^2-x_k^2)^2}~,\qquad  i,j,k\ {\rm cyclic}~.
\ee
Introducing the lowering and raising operators
$\xi_{\pm} :=\xi_1\pm i\xi_2~,$
the spin dependent part $H_{\rm spin}$ of the Hamiltonian (\ref{HHHs}) 
can be written in the form
\be
\label{Hrot}
H_{\rm spin} = {1\over 8}\left(\xi^2_+ + \xi^2_-\right)(V_1-V_2)
	      +{1\over 8}\left(\xi_+\xi_- +\xi_-\xi_+\right) (V_1+V_2)
              +{1\over 2}\xi^2_3 V_3~,
\ee
Since the Hamiltonian (\ref{HHHs}) commutes with $\vec{J}^2$ and $J_z$,
the energy eigenfunctions $\Psi_{JM}$ can be characterized by the two
quantum numbers $J$ and $M$.
Furthermore we shall expand the wave function $\Psi_{JM}$ in the basis of
the well-known $D$ functions \cite{Brink}, which are the common eigenstates
of the operators $\vec{J}^2=\vec{\xi}^2$, $J_z$ and $\xi_3$
with the eigenvalues $J,M$ and $k$ respectively,
\be
\Psi_{JM}(x_1,x_2,x_3;\alpha,\beta,\gamma)
=\sum_{k=-J}^J i^J\sqrt{{2J+1\over 8\pi^2}}
\Psi_{JMk}(x_1,x_2,x_3)D_{kM}^{(J)}(\alpha,\beta,\gamma)~,
\ee
where $(\alpha,\beta,\gamma)$ are the Euler angles.
We have the relations
\be
\xi_3D_{kM}^{(J)}=k D_{kM}^{(J)}\ ,\ \
\xi_{\pm} D_{kM}^{(J)}=\sqrt{(1\pm k +1)(1\mp k)} D_{k\pm 1\ M}^{(J)}~.
\ee
The task to find the spectrum of the Hamiltonian (\ref{HHHs})
then reduces to the following eigenvalue problem
for the expansion coefficients $\Psi_{JMk}$ for fixed values of $J$ and $M$
\be
\label{eigenvp}
\sum_{k=-J}^J \left[
\left(H_0-E\right) \delta_{k^\prime,k}
 +(-1)^J (2J+1)\int {\sin\beta d\alpha d\beta d\gamma \over 8\pi^2}
D_{k^\prime M}^{(J)\ast}(\alpha,\beta,\gamma)H_{\rm spin}
  D_{kM}^{(J)}(\alpha,\beta,\gamma)
\right] \Psi_{JMk} = 0~.
\ee

Since the spin part $H_{\rm spin}$ of the Hamiltonian does not commute
with $\xi_3$, nondiagonal terms arise, coupling different values of $k$.
We shall in the following limit ourselves to the case of spin-1.
Using the linear combinations \cite{Landau}
\bea
\Psi_{1}(x_1,x_2,x_3) &:=&
       {1\over \sqrt{2}}\left[\Psi_{J=1,M,k=1}(x_1,x_2,x_3)
-\Psi_{J=1,M,k=-1}(x_1,x_2,x_3)\right]~,\\
\Psi_{2}(x_1,x_2,x_3) &:=&
       {1\over \sqrt{2}}\left[\Psi_{J=1,M,k=1}(x_1,x_2,x_3)
+\Psi_{J=1,M,k=-1}(x_1,x_2,x_3)\right]~,\\
\Psi_{3}(x_1,x_2,x_3) &:=& \Psi_{J=1,M,k=0}(x_1,x_2,x_3)~,
\eea
the corresponding eigenvalue problem (\ref{eigenvp}) for spin-1 decouples
to the following three Schr\"odinger equations
for the wave functions $\Psi_a(x_1,x_2,x_3)$
\be
\label{effSch1}
\left[-{1\over 2}\sum_{i=1}^3{\partial^2\over\partial x_i^2}
+{g^2\over 2}\sum_{i<j} x_i^2 x_k^2
+ V^{\rm eff}_a(x_1,x_2,x_3)
\right]\Psi_a(x)=E\Psi_a(x)~,
\ \ \ \ a=1,2,3 ~,
\ee
with the effective potential
\be
\label{effSch2}
V^{\rm eff}_a(x_1,x_2,x_3):={1\over 2}(V_b+V_c)= {1\over 2}
\left({x_a^2+x_c^2\over (x_a^2-x_c^2)^2}+
{x_a^2+x_b^2\over (x_a^2-x_b^2)^2}\right)~,\qquad  a,b,c\ {\rm cyclic}~.
\ee
In the spin-1 sector we have therefore succeeded to reduce the
Schr\"odinger equation to three effective Schr\"odinger equations
for the scalar degrees of freedom with an additional effective potential
induced by the rotational degrees of freedom.
Since the effective potentials $V_i^{\rm eff}$ are related via
 cyclic permutation 
\be
\sigma_{123}V_1^{\rm eff}= V_2^{\rm eff}\sigma_{123}~,\ \ \ \
\sigma_{123}V_2^{\rm eff}= V_3^{\rm eff}\sigma_{123}~,\ \ \ \
\sigma_{123}V_3^{\rm eff}= V_1^{\rm eff}\sigma_{123}~,
\ee
all energy levels in the spin-1 sector are threefold degenerate.

As in the spin-0 sector we may use the variational approach to obtain
an upper bound for the lowest spin-1 state. 
The variational ansatz
\be
\Psi_a(x_1,x_2,x_3):= (x_a^2-x_b^2)(x_a^2-x_c^2)
\prod_{i=1}^3 \Psi_{0}(\omega_i,x_i)
\ee
satisfies both the boundary conditions (\ref{bc1})
and (\ref{bc2}) and vanishes at the
singularities of the additional effective spin-1 potential $V_{\rm eff}$.
For the optimal values 
\be
\omega_a=1.1814~ g^{2/3},\quad\quad  \omega_b=\omega_c=2.34945~g^{2/3}~, 
\ee
we obtain the energy minimum 
\be
\label{varres}
E_{\rm spin-1}=8.6044~ g^{2/3}~.
\ee
Analogous treatments of higher spin states can be carried out
correspondingly.
Using the linear combinations \cite{Landau}
\bea
\Psi_{J|k|}^{\pm}(x_1,x_2,x_3) &:=&
       {1\over \sqrt{2}}\left[\Psi_{J,M,k}(x_1,x_2,x_3)
\pm\Psi_{J,M,-k}(x_1,x_2,x_3)\right]~,\ \ \ \ \ k\ne 0~,\\
\Psi_{J0}(x_1,x_2,x_3) &:=& \Psi_{J,M,k=0}(x_1,x_2,x_3)~,
\eea
and noting that there are no transitions between the states 
$\Psi_{J|k|}^{\pm}$ with even and odd $k$, and with $+$ and $-$ index, 
the corresponding eigenvalue problem (\ref{eigenvp}) for spin-J decouples
into four separate Schr\"odinger eigenvalue problems.
For spin-2 one finds one cyclic triplet of degenerate eigensates and
two singlets under cyclic permutation, for spin-3 two cyclic triplets
each consisting of three degenerate states and one singlet, and so on.  
The corresponding reduction on the classical level
using the integrals of motion (\ref{cHI}) has been done in \cite{GKMP}.

We conclude this subsection by pointing out that our variational result
(\ref{varres}) shows that the higher spin states appear already at rather 
low energies
and therefore have to be taken into account in calculations of the
low energy spectrum of Yang-Mills theories.

\section{Calculation of the topological susceptibility}

The explicit evaluation of the Witten formula (\ref{WF0})
for the topological susceptibility allows us to 
check the consistency of the results for the low energy spectrum obtained 
in Section IV using the variational approach.

Using the groundstate $S_1$ in (\ref{groundstate}), obtained from
minimization of the energy functional in the singlet sector including
irreducible states with up to $5$ knots, and the expressions for
the matrix elements of $B^2$ in the basis of irreducible states 
given in Appendix C, we obtain 
\be
\label{WF01}
{d^2E_0(\theta)\over d\theta^2}\big|_{\theta =0}^{\rm contact}=
+ \langle 0|\left({\alpha_s\over 2\pi}\right)^2 B^2|0\rangle
=+0.0005117 ~ g^{14/3}~(+0.0005119~ g^{14/3})
\ee
for the contact term in the Witten formula.
The number in brackets gives the corresponding result for up to $4$ knots.

Since the $Q$-operator is a spin-0 operator
and symmetric under cyclic permutations,
the propagator term involves only the singlet states in the spin-0 sector.
Using the formula for the matrix elements of the topological operator
$Q$ stated in Appendix C and including the lowest fifteen (ten) excitations 
$S_2,\dots ,S_{16}$ ($S_2,\dots ,S_{11}$)
obtained approximately in the variational calculation 
as eigenvectors of the truncated Fock space including irreducible
singlet states up to $5$ knots ($4$ knots), we obtain \footnote{
Here the lowest six excitations $S_2,\dots ,S_7$ are found to give the  
contributions, $-103.3~\cdot 10^{-6}~g^{14/3}$ 
($-107.7~\cdot 10^{-6}~g^{14/3}$), 
$-201.6~\cdot 10^{-6}~g^{14/3}$ ($-205.3~\cdot 10^{-6}~g^{14/3}$), 
$-124.1~\cdot 10^{-6}~g^{14/3}$($-120.4~\cdot 10^{-6}~g^{14/3}$),
$-8.8~\cdot 10^{-6}~g^{14/3}$($-9.3~\cdot 10^{-6}~g^{14/3}$),
$-27.3~\cdot 10^{-6}~g^{14/3}$($-18.4~\cdot 10^{-6}~g^{14/3}$)
and $-0.16~\cdot 10^{-6}~g^{14/3}$ ($-4.1~\cdot 10^{-6}~g^{14/3}$)
respectively. The contributions from the remaining higher 
excitations $S_8,\dots ,S_{16}$ ($S_8,\dots ,S_{11}$ for up to $4$ knots) 
are of the order of $5\cdot 10^{-6}$ or less and form a series which is 
rapidly decreasing with the number of knots.}
\be
\label{WF02}
{d^2 E_0(\theta)\over d\theta^2}\big|_{\theta =0}^{\rm prop}=
-2\sum_n{|\langle 0|Q|n\rangle|^2\over E_n-E_0}=
-0.0004819~g^{14/3}~(-0.0004622~ g^{14/3})~.
\ee 
We see that the sum of the contact contribution (\ref{WF01})
and the propagator contribution (\ref{WF02}) seem to tend to zero
when extending the variational calculation to Fock states of higher
and higher number of knots. 
For comparison we point out that using the irreducible singlet states  
$\Psi_{000}^{(0)+},\Psi_{001}^{(0)+},\dots $ up to $5$ knots ($4$ knots)
in (\ref{typeI})-(\ref{typeIII}) directly,
instead of the eigenstates $S_1,S_2,\dots ,S_{16}$ 
($S_1,S_2,\dots ,S_{11}$), we get  
$-0.0005205~ g^{14/3}$ for the contact contribution (\ref{WF01}) and
$-0.0003808~ g^{14/3}$ ($-0.0003761~ g^{14/3}$) for the propagator 
contribution (\ref{WF02}).
We herewith find strong support that our variational results
are in accordance with vanishing topological susceptibility (\ref{WF0}). 

\section{Concluding remarks}

In this paper we have analysed the quantum mechanics of spatially
homogeneous gauge invariant $SU(2)$ gluon fields with theta angle.
We have reduced the eigenvalue problem of the Hamiltonian of this toy 
model for arbitrary theta angle to the corresponding problem with zero 
theta angle.
The groundstate, its energy eigenvalue, its magnetic and electric
properties, as well the corresponding value of the "gluon condensate"
and the lowest excitations have been obtained with high accuracy
using the variational approach. Furthermore it has been shown that
higher spin states become already relevant at rather low energy.
The groundstate energy has been found to be independent of the
theta angle by construction of the explicit transformation
relating the Hamiltonians with different theta parameter.
An explicit calculation of the Witten formula for the topological
susceptibility using our variational results for the
groundstate and the low lying excitations give strong support
for the independence of the groundstate energy of theta thus
indicating the consistence of our results.
We have found a continuous spectrum and the corresponding eigenstates
of the topological operator in this approximation and shown that its
characteristics coincide with the Euclidean self(anti-)dual zero-energy
solutions of the classical equations of motion. They are the
analogs of the instanton solutions, but do not correspond to
quantum tunneling between different vacua.
The generalization of these investigations to $SU(2)$ field theory
following \cite{KP3} is under present investigation.

\acknowledgments

We are grateful for discussions with S.A. Gogilidze, J. Hoppe, D. Mladenov, 
H. Nicolai, P. Schuck, M. Staudacher, A.N.Tavkhelidze and J. Wambach.
A.M.K.  would like to thank Prof. G.R\"opke
for kind hospitality at the MPG AG ''Theoretische Vielteilchenphysik''
Rostock where part of this work has been done
and the Deutsche Forschungsgemeinschaft for providing a
stipendium for the visit.
This work was supported also by the Russian Foundation for
Basic Research under
grant No. 98-01-00101 and by the Heisenberg-Landau program.
H.-P. P.  acknowledges support by the Deutsche Forschungsgemeinschaft
under grant No. RO 905/11-3.

\begin{appendix}

\section{Topological charge operator,
zero-energy solutions of the classical 
Euclidean equations of motion and tunneling 
amplitudes}

In this Appendix the solution of the eigenvalue problem 
for the topological charge operator is described and the relation 
between its characteristics and the Euclidean zero energy trajectories 
of the unconstrained  Hamiltonian (\ref{eq:PYM}) discussed. 
We shall also discuss the role of these  Euclidean zero energy solutions 
of the classical equations of motion to tunneling from one 
valley to another. 

\subsection{The eigenvalue problem for the $Q$- operator}

The eigenvalue problem for the topological charge operator 
\be
\label{Qeigen}
Q |\Psi(t)\big>_\lambda =
      \lambda |\Psi(t)\big>_\lambda
\ee
in the  Schr\"odinger representation reduces 
to the solution for the following linear partial differential equation
\bea
x_1x_2\frac{\partial}{\partial x_3}\Psi_\lambda(x_1, x_2, x_3)
+x_2x_3\frac{\partial}{\partial x_1}\Psi_\lambda(x_1, x_2, x_3)
+x_3x_1\frac{\partial}{\partial x_2}\Psi_\lambda(x_1, x_2, x_3)
=-i{8\pi^2\over g^3} \lambda \Psi_\lambda(x_1, x_2, x_3)\label{eve}~.
\eea

The conventional method  of characteristics
relates this problem to  the solution of the set of 
ordinary differential equations
\be
\label{4equs}
\frac{dx_1}{x_2x_3}= \frac{dx_2}{x_3x_1}=\frac{dx_3}{x_2x_1}
\ee
The integral curves corresponding to (\ref{4equs}) can  be written
in the form
\be
\label{iom}
I_1 =x_2^2-x_1^2\ ,\ \ \ I_2=x_3^2-x_1^2~.
\ee
These integral curves promt us 
with the introduction of the new adapted coordinates 
$(\zeta,\eta,\rho)$
\be
\label{newcoord}
\zeta:=x_1\,, \quad \eta :=x_2^2-x_1^2\,, \quad \rho :=x_3^2-x_1^2\,.
\ee
Such functions can be used 
as a suitable coordinates on the subset 
\be\label{domain}
0< x_1 < x_2 < x_3 < \infty 
\ee
of the whole configuration space $R^+_3$. 
The subset  (\ref{domain}) corresponds to the domain
$0< \zeta < \sqrt{\eta+\zeta^2} < \sqrt{\rho+ \zeta^2}$. 
Due to the symmetry of the $Q$ -operator under arbitrary permutations 
of the canonical pairs  $x_i, p_i$ the results  can be 
extended to the whole $ R^+_3$.

Writing the wave function in terms of new variables 
\be
\Psi_\lambda(x_1,x_2,x_3)=: W_\lambda(\zeta,\eta,\rho)
\ee
the partial differential equation (\ref{eve}) reduces to the following
ordinary differential equation
\be
\label{ode}
\sqrt{\zeta^2+\eta}\sqrt{\zeta^2+\rho}\frac{\partial}{\partial\zeta}
W_\lambda(\zeta,\eta,\rho)=-i\lambda 
{8\pi^2\over g^3}
W_\lambda(\zeta,\eta,\rho)~.
\ee
The general solution of this equation can be written in the form
\be
\label{gensol1}
W_\lambda(\zeta,\eta,\rho)=\Psi_0(\eta,\rho)
\exp[i\lambda\frac{8\pi^2}{ g^3 \sqrt{\rho}}
F\left(\arctan
\left({\zeta\over\sqrt{\eta}}\right),
{\sqrt{\rho -\eta\over \rho}}\right)]
\ee
with the arbitray function $\Psi_0(\eta,\rho)$ and the Jacobi elliptic
integrals $F(z,k)$ of the first kind \cite{Bateman}.
In terms of the original coordinates $(x_1,x_2,x_3)$
the eigenfuctions for the topological charge operator
in the sector $x_1 < x_2 < x_3$ therefore have the form
\be
\label{gensol2}
\Psi_\lambda(x_1, x_2, x_3)\ =\Psi_0(x_2^2-x_1^2,x_3^2-x_1^2)
\exp\left[i\lambda
\frac{8\pi^2}{ g^3 \sqrt{x_3^2-x_1^2}} 
F\left(\arctan\left({x_1\over\sqrt{x_2^2-x_1^2}}\right),
\sqrt{{x_3^2-x_2^2\over x_3^2-x_1^2}}\right)\right]~.
\ee
In the other sectors the corresponding wavefunction is obtained from
(\ref{gensol2}) by cyclic permutation. 
Note that the eigenfunctions (\ref{gensol2}), which constitute
the most general solution of the eigenvalue problem (\ref{Qeigen}) for the 
$Q$ operator, do not satisfy the 
boundary conditions (\ref{bc1}) and (\ref{bc2}) necessary for the
Hermiticity of the $Q$ operator. 

In the next paragraph we will show that the characteristics 
of the topological charge operator 
(\ref{4equs})   coincide with 
of the equations which determine the zero-energy solutions of the Euclidean
classical equations of motion.

\subsection{Euclidean zero energy trajectories in Yang-Mills mechanics}

The Euclidean action
\be
S^{\rm Eucl}= \int d\tau \left[{1\over 2}\left(\frac{dx}{d\tau}\right)^2 
 + V(x) \right]~,
\ee
is obtained from the corresponding action in Minkowski space
by inverting the  potential $ V(x) \longrightarrow  -V(x)~$.
In the one dimensional case the solutions of equation
\be
\frac{dx}{d\tau} =\pm \sqrt{2V(x)}~,
\ee
coorespond to trajectories with zero Euclidean energy
\be
E^{\rm Eucl}={1\over 2}\dot{x}^2 - V(x)~,
\ee
and the same time satisfies the classical Euclidean equations of 
motion
\be
\label{EEOM}
\frac{d^2x}{d\tau^2} = {dV\over dx}~.
\ee
Such type of trajectories play an important role in the description 
of quantum mechanical tunneling phenomena \cite{Coleman}. 
In the case that the potential $V(x)$ has at least
two local minima, say at $x=-a$ and $x=+a$, with $V=0$, the Euclidean
zero energy trajectories starting at $-a$ and ending at $a$ correspond
to quantum tunneling into the classically forbidden region.
The Euclidean action for these  classical $E^{Eucl}=0$ trajectories 
\be
S^{\rm Eucl}\big\vert_{E=0}= \int dt\left({dx\over dt}\right)^2 = 
\int_{-a}^a dx\sqrt{2V(x)}~,
\ee
determine in the semiclassical limit the WKB amplitude for a particle 
to tunnel from $x=-a$ to $x=+a$
\be
|T(E=0)| = \exp \left[ -{1\over \hbar}\int_{-a}^a dx\sqrt{2V(x)}
                 \right](1+O(\hbar))~.
\ee

The  potential of the unconstrained system considered in the
present article has three valleys.
The question is whether there exist the trajectories
corresponding to tunneling between the valleys. 
To answer the question let us rescale the coordinates 
$x_i \rightarrow g^{-1}x_i $ and write down the Euclidean action
of  the model in the form  
\be
S^{\rm Eucl}= \frac{1}{2g^2}
\int d\tau \sum_{cycl.}\left(\dot{x}_i^2 + x_j^2x_k^2\right)~,
\ee
The equations of motion then read  
\be \label{eqmot}
\ddot{x}_i= x_i(x_j^2+x_k^2) ;\qquad  i,j,k\ {\rm cyclic}~.
\ee
The class of trajectories with zero energy 
\be
E^{\rm Eucl}=\frac{1}{2g^2}\sum_{cycl.}\left(\dot{x}_i^2-x_j^2x_k^2\right)~,
\ee
can be chosen as the solutions of the following system 
of first order equations
\be \label{zeeq}
\dot{x}_1=\pm x_2x_3\ ,\ \ \ \ \dot{x}_2=\pm x_3x_1\ ,\ \ \ \
\dot{x}_3=\pm x_1x_2\ .\ \
\ee
If we fix one and the same sign on the r.h.s. 
of all equations (\ref{zeeq})
then they completely coincide  with 
the characteristic equations (\ref{4equs}) of the $Q$-operator.
Furthermore from (\ref{iom}) we see that the 
zero  energy Euclidean  solutions admit no trajectories from one
$V=0$ minimum to another and they have no relation 
to the quantum tunneling phenomena.

\subsection{$Q$- operator and selfdual states}

The commutator of the Hamiltonian $H_0$ of (\ref{H--0}) in the spin-0 sector 
and the topological charge $Q$
\be
[H_0,Q] = -i{g^3\over 4\pi^2}\left((p_1p_2x_3 + p_2p_3x_1 + p_3p_1x_2)
 -gx_1x_2x_3(x_1^2+x_2^2+x_3^2)\right)
\ee
vanishes only in the subspace of states $|\Psi>$ which satisfy
the Euclidean self(anti-)duality conditions
\be
\label{SD}
p_i|\Psi\rangle =\pm gx_jx_k |\Psi\rangle  ;\qquad  i,j,k\ {\rm cyclic}~,
\ee
which are the quantum analogs of the Euclidean $E=0$ 
constraints (\ref{zeeq}) discussed before.
Rewriting the Hamiltonian $H_0$ in (\ref{H--0}) in the form
\be
H_0={1\over 2}\sum_{i,j,k\ cycl.} \left(p_i^2+g^2x_j^2x_k^2\right)=
\sum_{i,j,k\ cycl.} \left(p_i\pm gx_jx_k\right)^2
\pm {8\pi\over g^2}Q ~.
\ee
we see that the Hamiltonian $H_0$ and the topological operator $Q$
 coincide
on the subspace of the Euclidean self(anti-)dual states (\ref{SD})
\be
H|\Psi\rangle =\mp {8\pi\over g^2}Q|\Psi\rangle ~.
\ee
Comparing this discussion with the corresponding original situation
$H=(1/2)(\Pi_i^{a2}+B_i^{a2})$ and
$Q=-(\alpha_S / 2\pi)\Pi_i^{a}B_i^{a}$ in terms of the
constrained fields $\Pi_i^a$ and $A_i^a$, where
$H_0=\mp (8\pi/ g^2)Q $ only in the Euclidean
self(-anti)dual case $\Pi_i^{a}=B_i^{a}$, we see that Eq. (\ref{SD})
correspond to the
unconstrained analogs of the self(anti-)dual configurations in Euclidean
space.
The question arises whether there are any self(anti-)dual states (\ref{SD})
which are both eigenstates of $Q$ and $H$.
The solution
\be
\Psi^{\pm}_{\rm SD}:= \exp[\mp igx_1x_2x_3]~,
\ee
of the self(anti-)duality conditions (\ref{SD}) in Euclidean space
is neither eigenfunction of $H_0$ nor of $Q$
\be
H_0\Psi^{\pm}_{\rm SD}=\mp {8\pi\over g^2} Q \Psi^{\pm}_{\rm SD}
= \pm 2V\Psi^{\pm}_{\rm SD}~,
\ee
The well-known exact nonnormalizable zero-energy solution of the 
spin-0 Schr\"odinger equation (\ref{H--0}), 
\be
\Psi_0^\pm = A\exp[\mp gx_1x_2x_3]~ ,
\ee
which differs from functional $\Psi_{\rm SD}$
up to a factor of $i$ in the exponent, actually satisfies the
self(-anti)duality conditions in Minkowski space
\be
-i{\partial\over \partial x_i}\Psi^{\pm}_0=
\pm i g x_jx_k \Psi^{\pm}_0 ;\qquad  i,j,k\ {\rm cyclic} ~,
\ee
(corresponding to $\Pi^a_i=\pm i B^a_i$), but 
is not an eigenfunction of the topological operator $Q$
\be
-{8\pi\over g^2}Q\Psi_0 =
\pm i g^2(x_1^2x_2^2+x_2^2x_3^2+x_3^2x_1^2)\Psi_0=
\pm 2iV(x_1,x_2,x_3)\Psi_0~.
\ee
Finally we point out that the (approximate) groundstate wave function 
(\ref{groundstate}) obtained in the variational approach is not 
self(anti-)dual.

\section{Lower bound for the spin-0 Hamiltonian $H_0$}

In this appendix we would like to derive a lower bound for the spin-0 
Hamiltonian $H_0$ in (\ref{H--0}) along the line of Ref. \cite{Simon}.
Using the boundary conditions (\ref{bc1}) and (\ref{bc2}) and 
based on the  well-known operator inequality 
for oscilator on positive half line 
\bea
-{\partial^2\over\partial x^2} + y^2x^2 \ge 3|y|~,\nn
\eea
it follows that
\be
\label{HH_0}
H_0\ge {1\over 4}\left(-\Delta + 3\sqrt{2}g(x_1+x_2+x_3)\right) 
=:{1\over 2} H^\prime~.   
\ee
Since the Hamiltonian $H^\prime$ is known \cite{Simon} 
to have a discret spectrum, this is true also for $H_0$.   
An important open question is at which energy the groundstate is.
The knowledge of the groundstate energy of $H^\prime$ in inequality 
(\ref{HH_0}) would provide a lower bound for the groundstate energy 
of $H_0$. Due to the additive structure of the potential term in 
$H^\prime$ one can make a separable ansatz for the solution of 
the corresponding eigenvalue problem. 
The energy of the lowest such separable $H^\prime$ eigenstate 
satisfying the above boundary conditions (\ref{bc1}) and 
(\ref{bc2}) is
\be 
\label{sepest}
E^{\prime}_{\rm sep} = 6|\xi_0|(3g/2)^{2/3}= 9.1924~ g^{2/3}~,
\ee
where $\xi_0 = -2.3381 $ is the first zero of the Airy function.
From the operator inequality (\ref{HH_0}) and 
the lower bound (\ref{sepest}) 
for separable solutions of $H^\prime$ we therefore obtain the lower 
bound of the energy functional for separable functions   
\be 
\label{sepestPsi}
{\cal E}[\Psi_{\rm sep}] 
\ge {1\over 2} E^{\prime}_{\rm sep}= 4.5962~ g^{2/3}~.
\ee
Finally we remark, as has been pointed out already in \cite{gymqm},
that an analogous variational calculation for $H^\prime$
shows that also the
groundstate energy of the Hamiltonian $H^\prime$ in Eq. (\ref{HH_0}) 
is lower than the value $E^\prime_{\rm sep}$ 
of (\ref{sepest}) for the lowest separable solution.

\section{Matrix elements}

For the evaluation of the energy functional, the calculation of 
the value of the "gluon condensate" of the groundstate,
as well as the propagator term in the
Witten formula we need the matrix elements 
of $E^2$,$B^2$ and $Q$ with respect to the irreducible Fock space states 
(\ref{typeI})-(\ref{typeIII}) built from the basic Fock space states
(\ref{bel}). 

\subsection{Basic matrix elements for Hamiltonian and topological charge}

The matrix elements of $E^2$ and $B^2$ with respect to the basic Fock space 
states (\ref{bel}) with ($\omega_1=\omega_2=\omega_3=3^{1/3}~g^{2/3}$) 
are given by
\be
\langle \Psi_{m_1; m_2; m_3}|E^2|\Psi_{n_1; n_2; n_3}\rangle  
 = 3^{1/3}g^{2/3}
\sum_{\rm cyclic}{\cal H}_{m_in_i}^-\delta_{m_jn_j}\delta_{m_kn_k}~, 
\ee
and
\be
\langle\Psi_{m_1; m_2; m_3}|B^2|\Psi_{n_1; n_2; n_3}\rangle = 
3^{1/3}g^{2/3}
\sum_{\rm cyclic} 
{1\over 3}\delta_{m_in_i}{\cal H}_{m_jn_j}^+ {\cal H}_{m_kn_k}^+~,
\ee
where
\be
{\cal H}_{mn}^\pm := \delta_{mn}(2n+3/2)
\ \pm\ \delta_{m(n+1)}\sqrt{n+3/2}\sqrt{n+1}
\ \pm\ \delta_{(m+1)n}\sqrt{n+1/2}\sqrt{n}~.
\ee

For the topological operator $Q$ we have
\be
\langle\Psi_{m_1; m_2; m_3}|Q|\Psi_{n_1; n_2; n_3}\rangle = 
{2ig^{8/3}\over \pi^{7/2}3^{1/6}}
\sum_{\rm cyclic} 
{\cal Q}_{m_in_i}^-{\cal Q}_{m_jn_j}^+{\cal Q}_{m_kn_k}^+~,
\ee
where
\bea
{\cal Q}_{mn}^+ &:=& {1\over 1- 4(m-n)^2}~
           {(-1)^{m+n}(2m+1)!!(2n+1)!!\over \sqrt{(2m+1)!(2n+1)!}}~,\\
{\cal Q}_{mn}^- &:=& {m-n\over 1- 4(m-n)^2}~
           {(-1)^{m+n}(2m+1)!!(2n+1)!!\over \sqrt{(2m+1)!(2n+1)!}}~.
\eea

\subsection{The irreducible matrix elements in terms of the basic ones}

For any operator $O$ invariant under the permutations $\sigma_{ij}$,
such as $E^2$, $B^2$, the Hamiltonian $H_0$
and the topological operator $Q$, 
\be
[O,\sigma_{ij}]=0~,
\ee
the matrix elements of the irreducible states
$\langle\Psi^{(k)\pm}_{M}|O|\Psi^{(k)\pm}_{N}\rangle$ 
of type I (\ref{typeI}),
type II (\ref{typeII}), and type III (\ref{typeIII})
can then be expressed in terms of the basic matrix elements
\be
{\cal M}^{m_1m_2m_3}_{n_1n_2n_3}:=
\langle\Psi_{m_1; m_2; m_3}|O|\Psi_{n_1; n_2; n_3}\rangle 
\ee
as follows. For the type I, II and III singlet states we have
\be
\langle\Psi^{(0)+}_{mmm}|O|\Psi^{(0)+}_{nnn}\rangle = 
{\cal M}^{mmm}_{nnn}~,
\ee
\be
\langle\Psi^{(0)+}_{mmr}|O|\Psi^{(0)+}_{nns}\rangle = 
{\cal M}^{mmr}_{nns}
+2{\cal M}^{mmr}_{nsn}
\ee
and
\bea 
\langle\Psi^{(0)+}_{m_1m_2m_3}|O|\Psi^{(0)+}_{n_1n_2n_3}\rangle 
&=&{\cal M}^{m_1m_2m_3}_{n_1n_2n_3}+{\cal M}^{m_1m_2m_3}_{n_3n_1n_2}
+{\cal M}^{m_1m_2m_3}_{n_2n_3n_1}\nonumber\\
&& +{\cal M}^{m_1m_2m_3}_{n_2n_1n_3}
+{\cal M}^{m_1m_2m_3}_{n_3n_2n_1}+{\cal M}^{m_1m_2m_3}_{n_1n_3n_2}
\eea
respectively. The transition elements between the 
type I, II and III singlets are
\be
\langle\Psi^{(0)+}_{mmm}|O|\Psi^{(0)+}_{nns}\rangle 
= \sqrt{3} {\cal M}^{mmm}_{nns} 
\ee
\be
\langle\Psi^{(0)+}_{mmm}|O|\Psi^{(0)+}_{n_1n_2n_3}\rangle 
= \sqrt{6} {\cal M}^{mmm}_{n_1n_2n_3}
\ee
and
\be
\langle\Psi^{(0)+}_{mmr}|O|\Psi^{(0)+}_{n_1n_2n_3}\rangle 
= \sqrt{2} {\cal M}^{mmr}_{n_1n_2n_3}  
+ \sqrt{2} {\cal M}^{mmr}_{n_3n_1n_2}
+ \sqrt{2} {\cal M}^{mmr}_{n_2n_3n_1}~. 
\ee
For the doublets, which exist only for type II and III states, we have
\be
\langle\Psi^{(1,2)+}_{mmr}|O|\Psi^{(1,2)+}_{nns}\rangle
 = {\cal M}^{mmr}_{nns}  -{\cal M}^{mmr}_{nsn}
\ee
and
\bea 
\langle\Psi^{(1,2)+}_{m_1m_2m_3}|O|\Psi^{(1,2)+}_{n_1n_2n_3}\rangle 
&=&{\cal M}^{m_1m_2m_3}_{n_1n_2n_3}-(1/2){\cal M}^{m_1m_2m_3}_{n_3n_1n_2}
- (1/2){\cal M}^{m_1m_2m_3}_{n_2n_3n_1}\nonumber\\
&& +{\cal M}^{m_1m_2m_3}_{n_2n_1n_3}
- (1/2){\cal M}^{m_1m_2m_3}_{n_3n_2n_1}
- (1/2){\cal M}^{m_1m_2m_3}_{n_1n_3n_2}
\eea
for the type III doublets. Their transition elements are
\be 
\langle\Psi^{(1,2)+}_{mmr}|O|\Psi^{(1,2)+}_{n_1n_2n_3}\rangle 
=\sqrt{2}{\cal M}^{mmr}_{n_1n_2n_3}
-\left({\cal M}^{mmr}_{n_3n_1n_2}
+{\cal M}^{mmr}_{n_2n_3n_1}\right)/\sqrt{2}~.\nonumber\\
\ee
For the axial singlets we have
\bea 
\langle\Psi^{(0)-}_{m_1m_2m_3}|O|\Psi^{(0)-}_{n_1n_2n_3}\rangle 
&=&{\cal M}^{m_1m_2m_3}_{n_1n_2n_3}+{\cal M}^{m_1m_2m_3}_{n_3n_1n_2}
+{\cal M}^{m_1m_2m_3}_{n_2n_3n_1}\nonumber\\
&& -{\cal M}^{m_1m_2m_3}_{n_2n_1n_3}
-{\cal M}^{m_1m_2m_3}_{n_3n_2n_1}-{\cal M}^{m_1m_2m_3}_{n_1n_3n_2}
~.\eea
For the axial doublets we have
\bea 
\langle\Psi^{(1,2)-}_{m_1m_2m_3}|O|\Psi^{(1,2)-}_{n_1n_2n_3}\rangle 
&=&{\cal M}^{m_1m_2m_3}_{n_1n_2n_3}-(1/2){\cal M}^{m_1m_2m_3}_{n_3n_1n_2}
- (1/2){\cal M}^{m_1m_2m_3}_{n_2n_3n_1}\nonumber\\
&& -{\cal M}^{m_1m_2m_3}_{n_2n_1n_3}
+ (1/2){\cal M}^{m_1m_2m_3}_{n_3n_2n_1}
+ (1/2){\cal M}^{m_1m_2m_3}_{n_1n_3n_2}
~.\eea

\end{appendix}

\end{document}